\documentclass[useAMS,usenatbib]{mnras}
\usepackage{amssymb,amsmath,epsfig}
\usepackage{algorithm,algpseudocode,enumitem}

\newcommand\Step{{\sf Step }}

\begin{document}

\title[Analysing Data via Local Two-Sample Tests]{Local Two-Sample Testing: A New Tool for Analysing High-Dimensional Astronomical Data}
\author[P. E. Freeman et al.]{P.~E.~Freeman,\thanks{E-mail: pfreeman@cmu.edu} I.~Kim and A.~B.~Lee\\
Department of Statistics, Carnegie Mellon University, 5000 Forbes Avenue, Pittsburgh, PA 15213, USA\\
}

\date{14 July 2017}

\pagerange{\pageref{firstpage}--\pageref{lastpage}} \pubyear{2017}

\maketitle

\label{firstpage}

\begin{abstract}
Modern surveys have provided the astronomical community with a flood of
high-dimensional data, but analyses of these data often occur after their
projection to lower-dimensional spaces. In this work, we introduce a local
two-sample hypothesis test framework that an analyst may directly apply to 
data in their native space. In this framework, the analyst defines two classes
based on a response variable of interest (e.g.~higher-mass galaxies versus
lower-mass galaxies) and determines at arbitrary points in predictor space 
whether the local proportions of objects that belong to the two classes 
significantly differs from the global proportion. 

Our framework has a 
potential myriad of uses throughout astronomy; here, we demonstrate its
efficacy by applying it to a sample of 2487 $i$-band-selected galaxies
observed by the HST ACS in four of the CANDELS program fields. For
each galaxy, we have seven morphological summary statistics along with an
estimated stellar mass and star-formation rate. We perform two studies:
one in which we determine regions of the seven-dimensional space of 
morphological statistics where high-mass galaxies are significantly more
numerous than low-mass galaxies, and vice-versa, and another study where
we use SFR in place of mass. We find that we are able to identify such
regions, and show how high-mass/low-SFR regions are associated with
concentrated and undisturbed galaxies while galaxies in 
low-mass/high-SFR regions appear more extended and/or disturbed than
their high-mass/low-SFR counterparts.
\end{abstract}

\begin{keywords}
galaxies: evolution -- galaxies: high-redshift -- galaxies: statistics -- galaxies: structure -- methods: statistical -- methods: data analysis
\end{keywords}

\section{Introduction}

Modern astronomical data are intrinsically high-dimensional; for any
given object, we may have images and photometric magnitudes (and perhaps 
spectra), as well as estimates of mass, star-formation rate, metallicity, etc.
Astronomical data analysis, however, still often operates in low
dimension, due less to a lack of will than to a lack of tools that 
astronomers can wield to effectively analyse data in their native spaces.

One specific area in which the ability to work with high-dimensional data is useful
is in the analysis of galaxy morphology. Morphological studies are key to 
understanding the evolutionary histories of galaxies and to constraining 
theories of hierarchical 
structure formation. (For a recent review, see, e.g.~\citealt{Conselice14}.)
A galaxy's morphology indicates
its current state (is it undergoing a merger? is it compact and quiescent?)
and may contain information about its assembly history (is it undergoing
a post-merger burst of star formation? does its central bulge indicate
past mergers?). 
We may think of a galaxy's morphology
as a continuous surface brightness function observed in three
dimensions (two spatial, one wavelength) that is
sampled from a distribution of such functions. Due to finite resolution,
what we actually observe is a pixelated and discretized version 
of the sampled function. Discretization helps us by moving
morphological analysis from the realm of infinite dimensionality\footnote{
In practice, we would need an infinite number of parameters to fully 
model surface brightness; S\'ersic profiles (\citealt{Sersic63}), 
for instance, are insufficient.}
to the finite realm, but the dimensionality (i.e., the number of image pixels
times the number of wavelengths at which image data are collected) is
still very large and thus analyses are still subject to the ``curse
of dimensionality."
To make analyses tractable, one conventionally reduces
the dimensionality further by, e.g.~computing summary statistics, 
which may be either parametric
(e.g.~the S\'ersic index $n$; \citealt{Sersic63}) or
nonparametric (e.g.~the Gini coefficient $G$; \citealt{Abraham03},
\citealt{Lotz04}).

Let $T$ represent a collection of morphological statistics.
We may model an ensemble of galaxies by a sample from a distribution
of moderate or high dimensionality
\begin{equation}
f(T;z,\lambda_{\rm obs};M_\ast,S,... \vert \theta) \,,
\end{equation}
where $z$ and $\lambda_{\rm obs}$ are redshift and observed wavelength,
$M_\ast$ and $S$ are galaxy stellar mass and star-formation
rate, and $\theta$ collectively represents the cosmological and
astrophysical parameters that govern structure formation.
A goal that is potentially realizable in the near future is to
statistically infer $\theta$, by comparing samples from estimates
$\widehat{f}$ derived from astronomical surveys with samples from
simulation models of $f(T;z,\lambda_{\rm obs};M_\ast,S,... \vert \theta)$ 
(e.g.~from the
{\em Illustris} and {\em Eagle} projects; \citealt{Vogelsberger14}, 
\citealt{Schaye15}). 
For example, likelihood-free methods, such as Approximate Bayesian
Computation (ABC; e.g.~\citealt{Weyant13} and references therein), 
currently rely on comparing a few
derived summary statistics instead of comparing two samples directly
in higher dimensions. In the meantime,
many authors attempt to infer the relationship between parametric
and/or nonparametric structure statistics and other statistics of interest:
$M_\ast$ and $S$, and in particular the ``main sequence" on the 
$M_\ast$-$S$ diagram
(e.g.~\citealt{Wuyts11}, \citealt{Elbaz11}, \citealt{Salmi12},
\citealt{Barro14}, \citealt{Brennan17}; see also \citealt{Snyder15} for
a similar analyis of simulated {\em Illustris} galaxies); 
the fraction of quenched galaxies
(e.g.~\citealt{Lang14}, \citealt{Bluck14}, \citealt{Woo15}, \citealt{Peth16},
\citealt{Bluck16}, \citealt{Teimoorinia16}; see also 
\citealt{Huertas16}); rest-frame colour (e.g.~\citealt{Wake12}); and
local environment (e.g.~\citealt{Lackner13}).\footnote{
Note that such inference
stands in constrast to using structure statistics to {\em predict}
classification; e.g.~\citealt{Simmons17} and references therein.}
(See also \citealt{Bell12}, who compare morphological statistics of 
star-forming and 
quiescent galaxies over cosmic time, and \citealt{Fang15}, who apply
unsupervised learning methods to structure statistics.)
However, save one exception which we mention below, all of these
authors work with one or two morphological statistics 
at a time instead of working
with an entire ensemble (which may include the effective radius, the
axis ratio, and the S\'ersic index, in addition to statistics
introduced via bulge-disc decompositions, the Gini and $M_{20}$ statistics,
etc.).\footnote{
\cite{Teimoorinia16} may be also be considered an exception,
in that they apply an artificial-neural-network algorithm that relates
eight statistics to quenching fraction, but ultimately their interest
lies in determining which subsets of two or three statistics have
the greatest predictive power.}
By concentrating their
efforts on projections of the ensemble rather than the full ensemble itself, 
the authors cannot truly map out dependencies between variables.

In this work, we present a new statistical framework which utilizes
local two-sample hypothesis tests. Astronomers will find this framework
useful for 
detecting and quantifying dependencies within statistical spaces of
{\em moderate or high} dimension. 
In particular, local two-sample tests can identify objects 
that lie in regions of 
predictor space where the estimated proportion of a particular defined 
class of objects (e.g.~galaxies of high mass, or of low metallicity, etc.) 
differs significantly from the global proportion.
There are a myriad of applications for this framework;
we demonstrate it by exploring 
the relationship between nonparametric structure statistics$-$namely
the seven image summary statistics $M,I,D$ (\citealt{Freeman13}),
$G,M_{20}$ (\citealt{Abraham03}, \citealt{Lotz04}), and $C,A$ 
(\citealt{Abraham94}, \citealt{Abraham96a}, \citealt{Abraham96b},
\citealt{Conselice03})$-$and
estimated stellar mass $M_\ast$ and star-formation rate $S$.
We note that our work has superficial
similarity to that done by \cite{Peth16}, who study the relationship
between the same seven image statistics and stellar quenching. However, their
work utilizes clustering$-$the authors first determine principal 
components for the seven statistics, then
use agglomerative hierarchical clustering to identify ten galaxy groups
(plus outliers) within PC space.
We on the other hand divide individual galaxies into groups based on 
defined response variables (estimated stellar mass, etc.), and then
identify locally significant differences between the two populations
{\em without} pre-clustering the data.

In Sec.~\ref{sect:methods}, we outline our local 
two-sample hypothesis test framework.\footnote{
The interested reader may find
{\tt R} functions for carrying out local two-sample tests 
at {\tt github.com/pefreeman/ltst}.}
(For more detail, see Kim \& Lee, in preparation.)
In Sec.~\ref{sect:apply}, we demonstrate its
efficacy by applying it to a sample
of 2487 $i$-band-selected (rest-frame wavelength $\approx$4{,}500 \AA)
galaxies observed by the {\em Hubble} Space Telescope's Advanced Camera 
for Surveys. We perform two studies: a morphology-mass study, where
we identify galaxies that lie in predominantly high-mass or low-mass
regions (or neither), and one where the division into two samples is based on 
star-formation rate rather than mass. In Sec.~\ref{sect:summary} we
summarize our findings.

\section{Local two-sample hypothesis testing}

\label{sect:methods}

Suppose that we are
given a set of $n$ measurements of an astronomical object, and that
our interest lies in determining those regions of the $n$-dimensional 
space of statistics where objects of particular class, class $y^0$ 
(perhaps defined as belonging to a particular redshift bin, or a particular
range of masses, etc.), are observed in greater or lesser proportions than
the class's global proportion.
One way to identify these regions is via
the use of two-sample, or homogeneity, tests. Let $P^0$ and $P^1$ 
represent distributions from which the independent samples
$x_1^0,\cdots,x_n^0$ and $x_1^1,\cdots,x_m^1$ are drawn, respectively.

In order to define regions where class proportions differ from their
global values, one needs to utilize a {\em local} two-sample test.
In a global two-sample test, we would compare the hypotheses
\begin{equation}
H_0~:~P^0 = P^1 ~~ {\rm against} ~~ H_1~:~P^0 \neq P^1 \,.
\end{equation}
Examples of such tests include
the maximum mean discrepancy test (MMD; \citealt{Gretton12}) and
the energy distance test (\citealt{Szekely04}), both being nonparametric
extensions of classical $t$ tests.
However, such tests can only
provide binary results: ``reject the null hypothesis" or 
``fail to reject the null hypothesis." In the current context,
rejection of the null hypothesis is typically uninteresting;
what is interesting is determining {\em where} in the
space of galaxy morphological statistics the distributions $P^0$ and $P^1$ 
significantly differ. Hence the need for local testing.

\begin{algorithm*}[ht]
\caption{Local Two-Sample Testing via Asymptotic Normality}
\begin{algorithmic}
\State \textbf{Input:} iid training samples $\{x_{i,0}\}_{i=1}^{n}$ and $\{x_{i,1}\}_{i=1}^{m}$.\\
\hspace{0.4in} Defined test points $\{x_1,\ldots,x_t\}$.
\begin{itemize}[leftmargin=1.5cm]
\item[\Step 1.] Fit random forest regression to training samples $\{x_{i,0}\}_{i=1}^{n}$ and $\{x_{i,1}\}_{i=1}^{m}$.
\item[\Step 2.] Calculate the test statistic $T(x_i)$ at the $t$ test points.
\item[\Step 3.] Compute the $p$-value at each test point: $p_i = 2\Phi(-\vert T(x_i) \vert)$, where $\Phi(\cdot)$ is the cumulative distribution function for the standard normal distribution.
\item[\Step 4.] Apply the Benjamini-Hochberg procedure to correct for the
number of tests (\citealt{Benjamini95}). Let $\{x_{(1)},\ldots,x_{(t)}\}$ be those data, sorted
by ascending (adjusted) $p$-value. Let $y$ be the identification of one of the
two discrete classes (e.g., high mass, in a high mass-low mass comparison). 
For each datum, conclude
  \begin{itemize}
  \item $\mathbb{P}(Y=y \vert x_{(i)}) > \mathbb{P}(Y=y)$ if $T_{(i)} > 0$ and the adjusted $p$-value is $p_{(i)} < p_{\rm thr}$
  \item $\mathbb{P}(Y=y \vert x_{(i)}) < \mathbb{P}(Y=y)$ if $T_{(i)} < 0$ and the adjusted $p$-value is $p_{(i)} < p_{\rm thr}$
  \end{itemize}
\end{itemize}
\hspace{0.54in} (The threshold $p$-value for rejecting the null hypothesis, $p_{\rm thr}$, is typically 0.05 but may be adjusted.)\\
\hspace{0.54in} For all other points, conclude $\mathbb{P}(Y=y \vert x_i) = \mathbb{P}(Y=y)$.
\end{algorithmic}
\label{alg:twosample}
\end{algorithm*}

In the flow cytometry literature, \cite{Roederer01} address the problem
of detecting differences between two samples in a multi-dimensional space.
Their method partitions the space into hypercubes, and 
identifies those hypercubes where $P^0 \neq P^1$. To capture
detailed local structures, it is natural to shrink the volume of each 
hypercube as the overall sample size increases, eventually approaching a
point-wise test in the limit of large sample sizes. We thus propose a
point-wise tests for differences at specified points ($x_1,\ldots,x_p$):
\begin{eqnarray}
H_{i,0}~&:&~\mathbb{P}(Y=y \vert X = x_i) = \mathbb{P}(Y = y) ~~ {\rm against} \nonumber \\
H_{i,1}~&:&~\mathbb{P}(Y = y \vert X = x_i) \neq \mathbb{P}(Y = y) \,.
\end{eqnarray}
Such a test is equivalent to testing for differences in density at the
specified points. Unlike \cite{Duong13}, who uses kernel density estimation
to find locally significant differences between two samples, we also
propose combining point-wise testing with a supervised learning method (such
as regression) that does not rely on estimating densities. Our proposed test
statistic is
\begin{equation}
T(x_i) = \widehat{\mathbb{P}}(Y = y \vert X = x_i) - \widehat{\mathbb{P}}(Y = y) \,,
\label{eqn:ts}
\end{equation}
where the estimated class prior
$\widehat{\mathbb{P}}(Y=y)$ is the fraction of observed objects of
class $Y = y$.

A challenge is to estimate the
``class posteriors" $\mathbb{P}(Y = y \vert X = x)$.
A principal advantage of our local two-sample testing framework is that we can 
take advantage of many existing regression methods for multi-dimensional
data. By choosing a suitable regression method, we can adapt to different
types of structure in the data as well as to different types of data, 
such that our test can potentially achieve high statistical power.
In this work, we apply random forest regression to estimate the class
posteriors. 
One advantage of random forests is that one can easily work with different
types of predictors, and unlike kernel smoothers, one does not need to 
specify a distance metric in the predictor space. 
Also, it performs de facto variable selection by providing measures
of variable importance (Fig.~\ref{fig:varimp}): not only can we identify
locally significant regions in the predictor space and {\em how} two
populations differ, we can also identify {\em which} summary statistics
are the most important in distinguishing the two populations.
And yet one more advantage of applying random forests to our data
lies in the work of \cite{Wager15},
who describe a random forest variant that yields predictions
that are both asymptotically unbiased and normally distributed under the 
null. We amend the test statistic given in eq.~(\ref{eqn:ts}):
\begin{eqnarray}
T(x_i) &=& \frac{\widehat{\mathbb{P}}(Y=1 \vert x_i) - \widehat{\mathbb{P}}(Y=1)}{\sqrt{\widehat{V}(x_i)}} \,, \label{eqn:test}
\end{eqnarray}
which under the null hypothesis converges to a standard normal distribution.
$\widehat{V}(x)$ is a consistent estimator of the variance of 
the random forest predictions 
based on the infinitesimal jackknife (\citealt{Wager14}).

Algorithm \ref{alg:twosample} shows the steps that we follow in our
analyses of galaxy data. Because the dimensionality of the predictor space
precludes us from defining a dense rectangular grid of points at which to
run local two-sample tests, we split our galaxy data into training
and test sets. We use the former to grow the forest, and we compute
two-sample test $p$ values using the latter. Given those $p$-values and
a significance criterion that is adjusted via the Benjamini-Hochberg 
false discovery rate algorithm with $\alpha = 0.05$ (\citealt{Benjamini95}), 
we determine whether each point lies in a region 
where the proportion of class $y$ galaxies is consistent with, or
significantly different than, the global proportion. (We choose $\alpha$ = 
0.05 as this is the standard value in statistical analyses but note that one
can adjust this value as necessary.)

\section{Application to galaxy data}

\label{sect:apply}

\subsection{Data}

\label{sect:data}

We demonstrate the efficacy of our 
framework by applying it to the analysis of
HST ACS $i$-band images from four fields
observed as part of the Cosmic Assembly
Near-IR Deep Extragalactic Legacy Survey (CANDELS; \citealt{Grogin11},
\citealt{Koekemoer11}).

\begin{figure}
  \centering
  \includegraphics[width=2.75in]{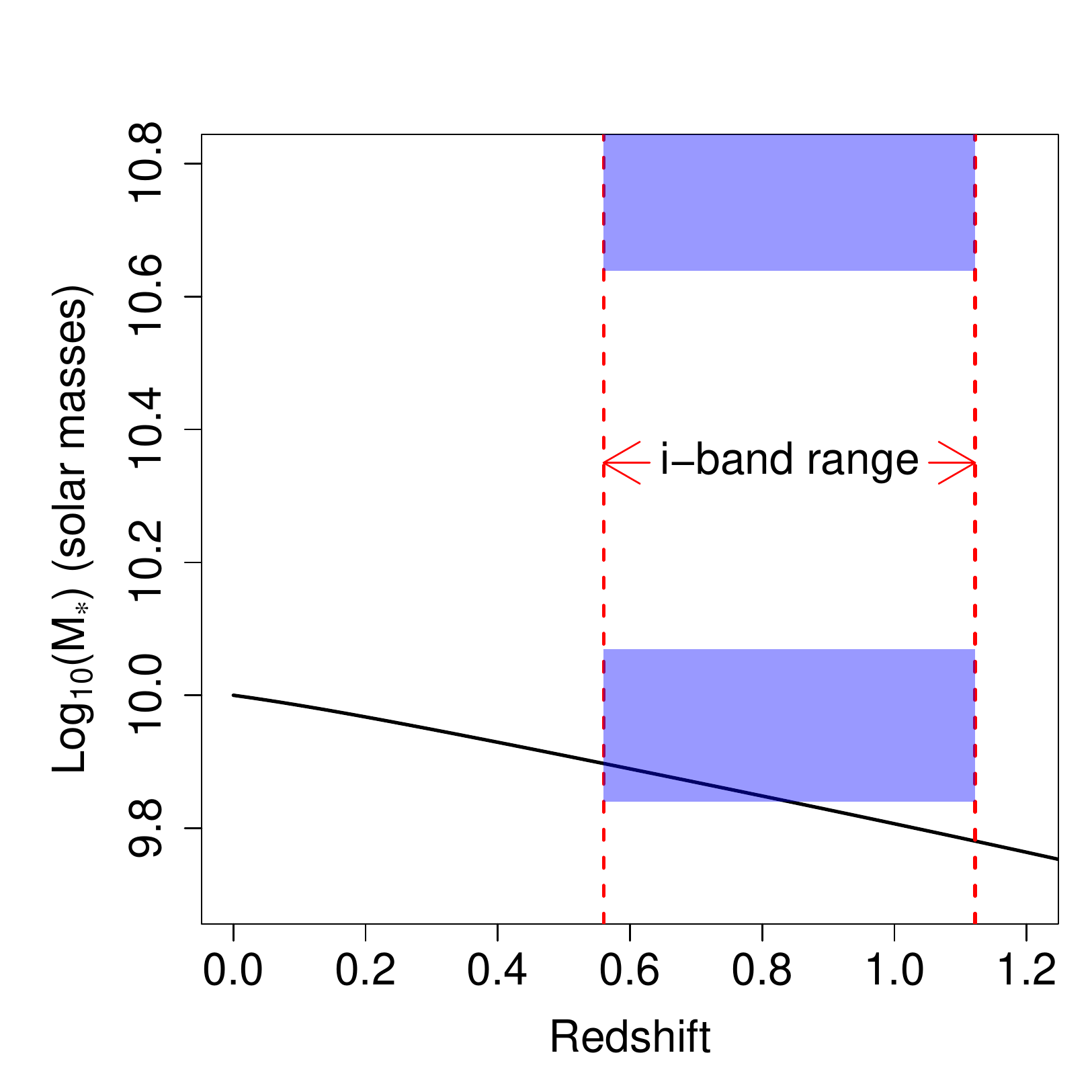}
  \caption{Minimum galaxy mass threshold for inclusion in our data sample
           as a function of redshift (black solid line), as estimated via
           the algorithm of Behroozi, Wechsler, \& Conroy (2013). In their
           scheme, if one evolves a sample of galaxies whose masses are all the
           threshold mass for a particular redshift from that redshift
           to the present day, the present-day median mass among those 
           galaxies would be $\log_{10}(M_\ast) \approx 10$.
           The red dashed lines indicate the
           redshift boundaries for our data sample: $z \in [0.56,1.122]$.
           We adopt as a minimum mass the average threshold over this 
           range: $\log_{10}(M_{\rm min}) = 9.84$. The lower and upper blue
           rectangles indicate the redshift-mass limits for our low-mass
           and high-mass quartile samples, respectively. (The upper 
           rectangle is artificially truncated by the plot boundary; in 
           actuality it extends to $\log_{10}(M_{\ast}) \approx 12.4$.)
          }
  \label{fig:behroozi}
\end{figure}

To construct our data sample, we begin
by defining a range of redshifts such that rest-frame 4500 \AA~is 
observed within the HST ACS F814W filter (i.e.~within the $i$-band).
We adopt the knee at $\approx$9550 \AA~in the filter transmission curve
as our upper wavelength bound, with matching lower bound at $\approx$7020 \AA;
thus $z \in [(7020-4500)/4500,(9550-4500)/4500] = [0.560,1.122]$.
Next, we apply magnitude, mass, and redshift cuts to the full four-field
CANDELS galaxy sample. The CANDELS Team Release mass catalogs 
include estimates of $H$-band magnitudes and spectroscopic redshifts
(if available) for each galaxy, as well as the 
median of a number of stellar mass estimates ($\widehat{M}_{\rm med}$; 
see \citealt{Mobasher15} and \citealt{Santini15}). For those galaxies
lacking spectroscopic redshifts, we utilize a photometric redshift 
conditional density estimate ${\widehat p}(z)$ made using a hierarchical
Bayesian technique that combines the output of five separate photometric 
redshift estimators summarized
in \cite{Dahlen13} (D.~Kodra \& J.~Newman, private communication). To
be included in our sample, a galaxy must have
\begin{itemize}
\item an $H$-band magnitude $H < 25$;
\item a spectroscopic redshift $z_{\rm spec} \in [0.560,1.122]$, or, if
the spectroscopic redshift is not available, an integrated CDE
$\int_{0.560}^{1.122} {\widehat p}(z) dz$ that is $\geq 0.8$; and
\item a mass $\widehat{M}_{\rm med} \geq 10^{9.84} M_\odot$.
\end{itemize}
We assume a mass threshold of $10^{10}$ M$_{\odot}$ at $z = 0$ 
and use the algorithm of \cite{Behroozi13} to adjust it downwards as
$z$ increases so as to maintain an approximately 
constant comoving galaxy number density (see Fig.~\ref{fig:behroozi}).
The average of the threshold curve over the range $z = [0.560,1.122]$
is $10^{9.84} M_\odot$. (We use the average of the curve rather than
the curve itself so as to ensure that the distribution of redshifts
in the low-mass and high-mass quartile samples that we analyse below
are similar.)

\begin{center}
\begin{table}
\centering
\caption{Sample size by CANDELS field.}
\begin{tabular}{cccc}
\hline
Field & Total & F814W-selected \\
\hline
COSMOS & 38\,671 & 704 \\
EGS    & 41\,457 & 539 \\
GOODSN & 35\,451 & 785 \\
UDS    & 35\,932 & 459 \\
\hline
Total  & 186\,441& 2487 \\
\hline
\label{tab:data}
\end{tabular}
\end{table}
\end{center}

\newpage

Our final data sample consists of
2487 galaxies, of which 891 have measured spectroscopic redshifts.
Image summary statistics$-$namely, $M,I,D$ (\citealt{Freeman13}), 
$G,M_{20}$ (\citealt{Abraham03}, \citealt{Lotz04}), and $C,A$ 
(\citealt{Abraham94}, \citealt{Abraham96a}, \citealt{Abraham96b},
\citealt{Conselice03})$-$are
determined for each galaxy using our own {\tt R} software suite.\footnote{
\tt https://github.com/pefreeman/galmorph} 
(Note that our current definition of the multimode statistic $M$
differs slightly from that of \citealt{Freeman13}: 
we divide the variable $R_l$ shown in their equation 1 by the 
number of pixels in the segmentation map, which places a hard upper limit 
of 0.5 on the area ratio $R_{l,max}$, achieved when $A_{l,(1)} = A_{l,(2)} =
n_{\rm seg}/2$.)
We adopt the cataloged star-formation rates estimated by A.~Fontana 
(via ``method 6.C," described in \citealt{Mobasher15}).

For the analyses below, we split the data into training and test sets 
of 1787 and 700 galaxies, respectively. 
We then assign the smallest and largest
25\% of mass (or SFR) values for the training set galaxies
to the low-mass (or low-SFR) and high-mass (or high-SFR) groups, 
respectively.

\begin{figure}
  \centering
  \includegraphics[width=2.75in]{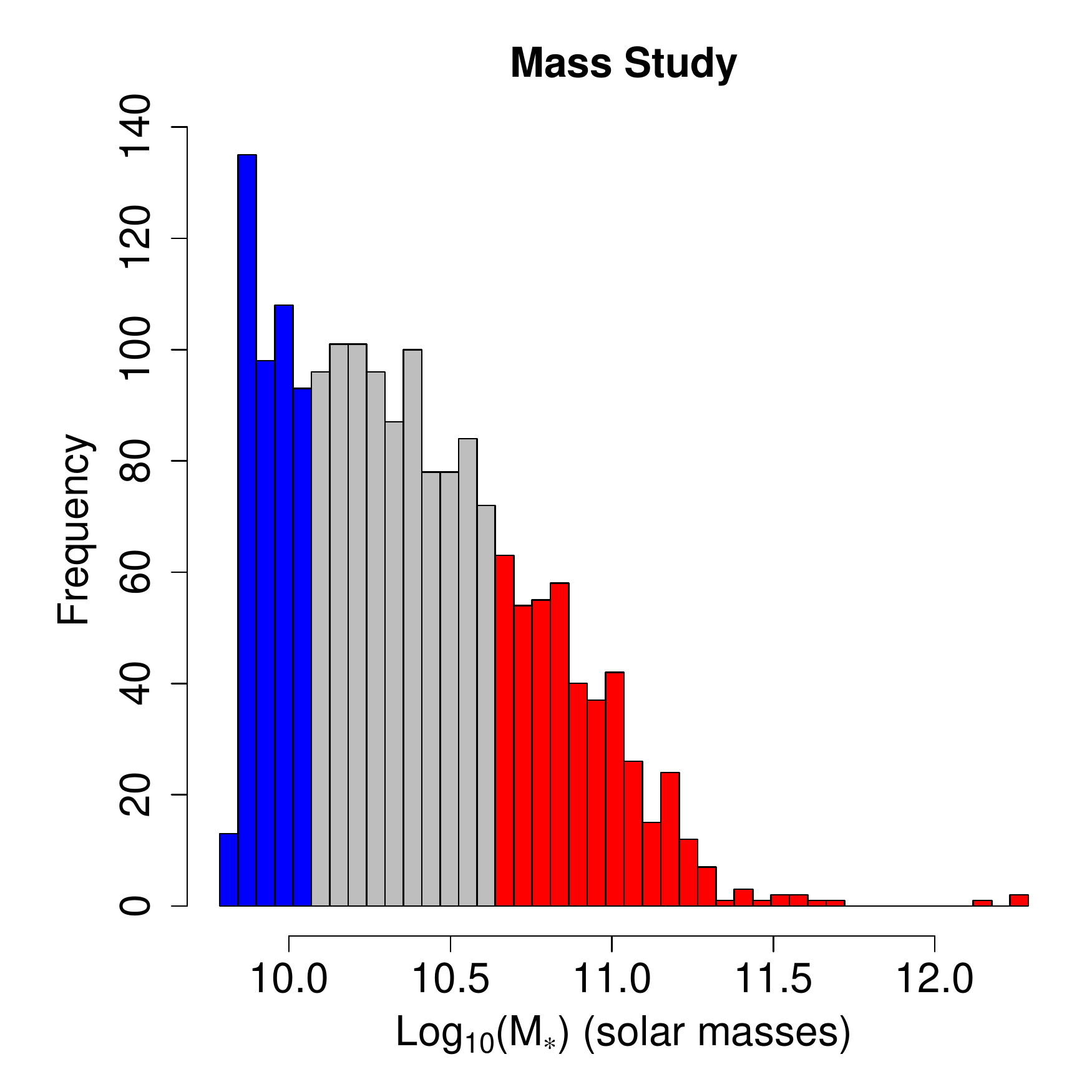}
  \includegraphics[width=2.75in]{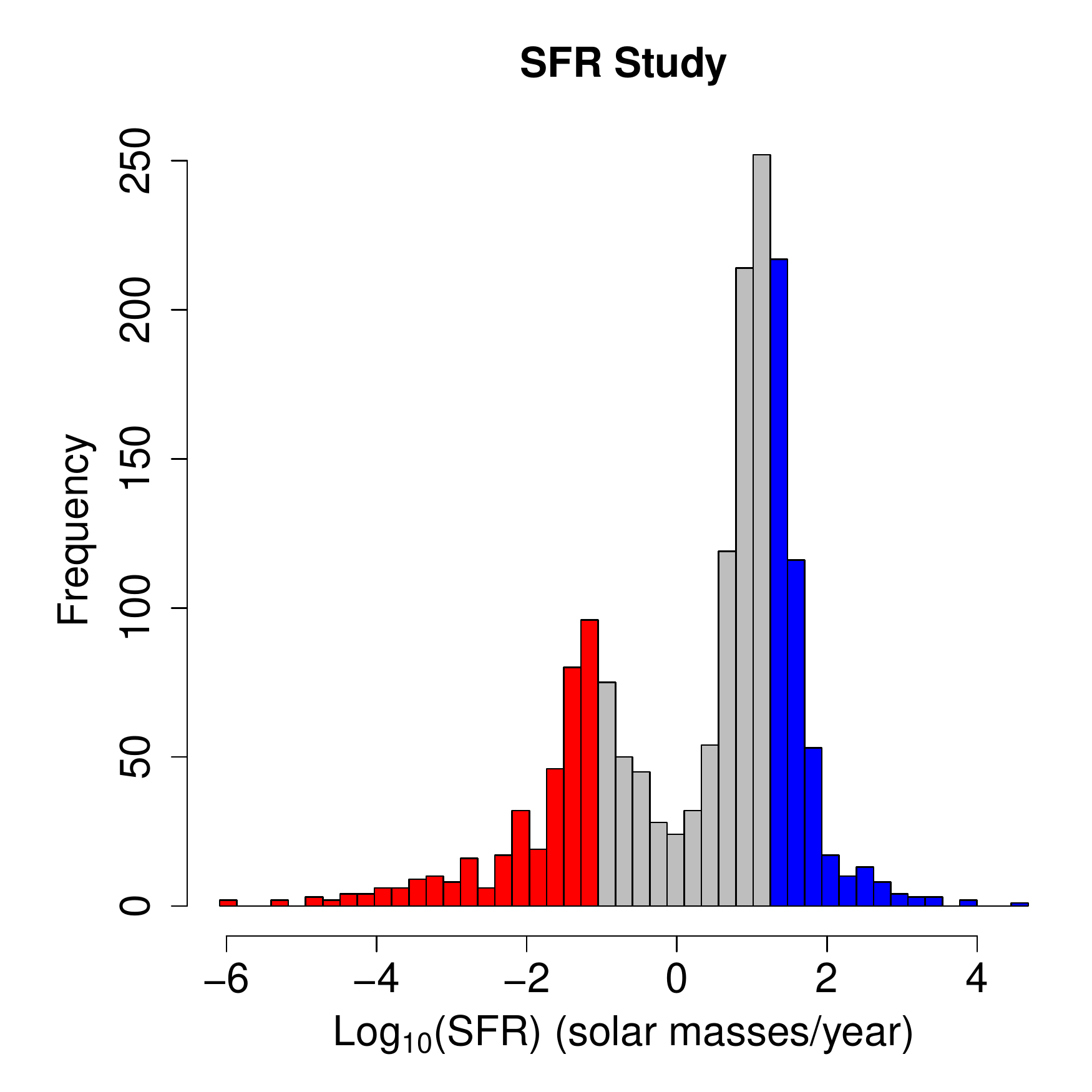}
  \caption{Distributions of masses (top) and star-formation rates (bottom)
           for the 1787 galaxies in the training set, with the lower and upper 
           quartiles
           highlighted. Note that the bottom histogram does not include
           79 galaxies for which $\widehat{S} = 0$ (but which are
           included in the low-SFR group).}
  \label{fig:dists}
\end{figure}

\begin{figure}
  \centering
  \includegraphics[width=2.75in]{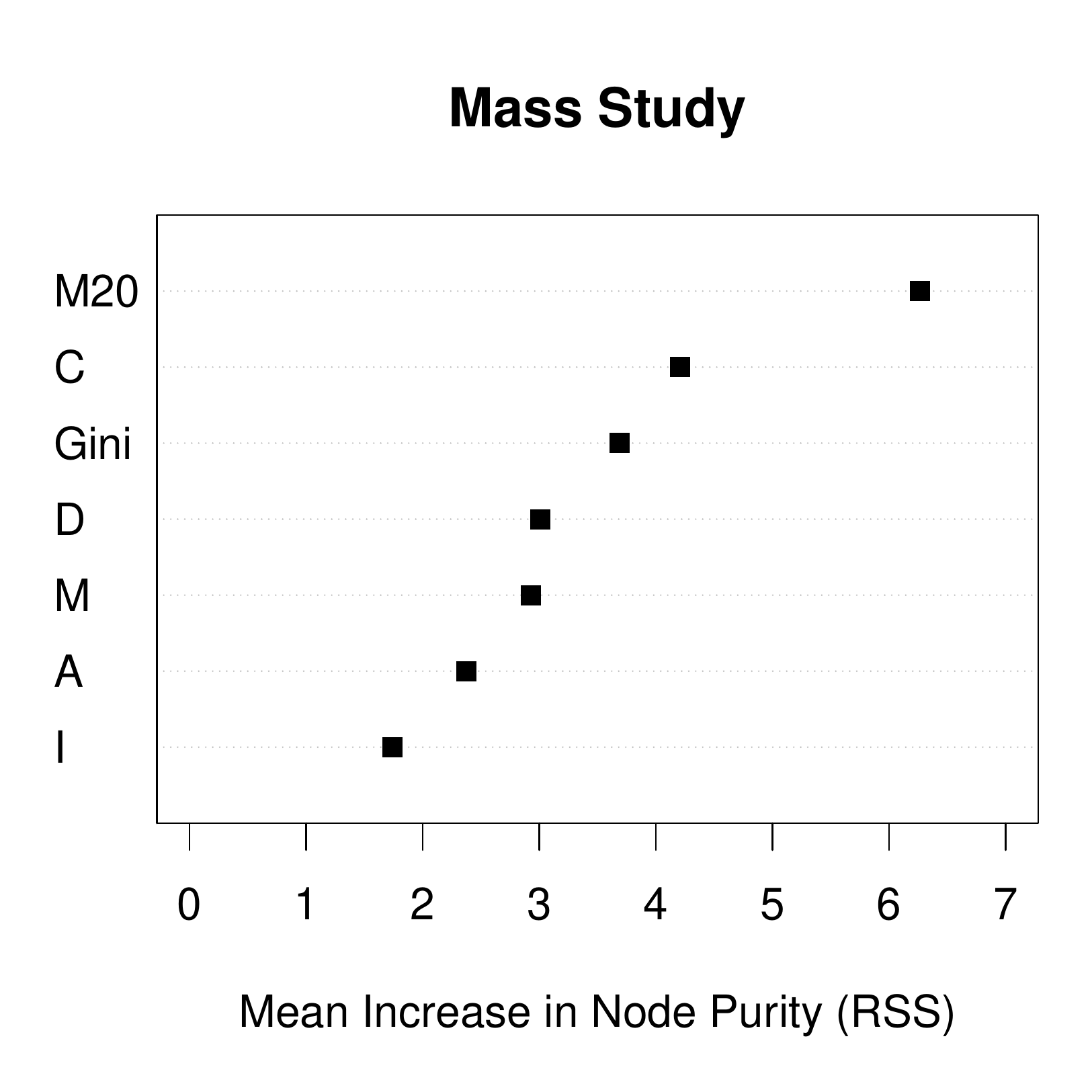}
  \includegraphics[width=2.75in]{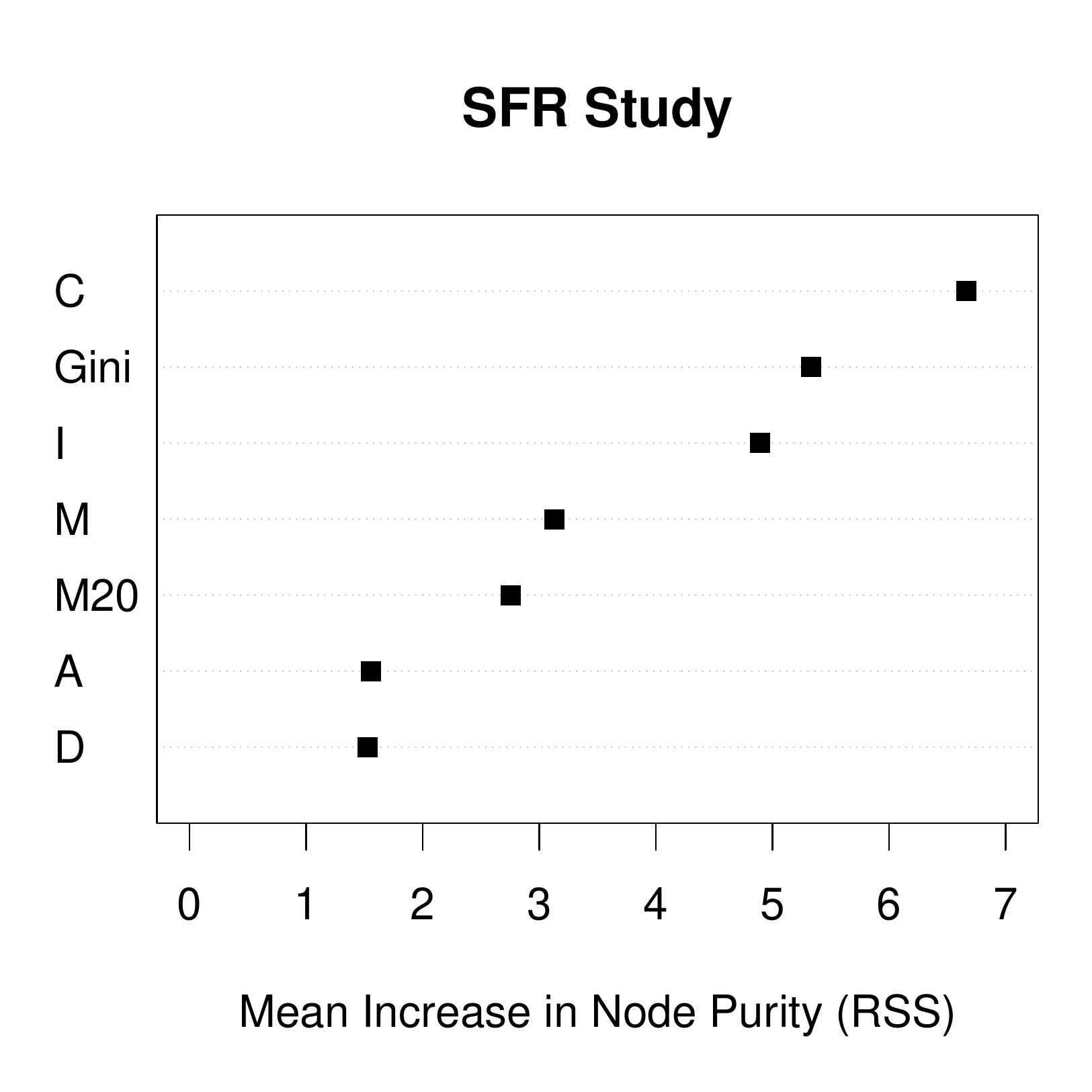}
  \caption{Variable importance measures for random forest regression, as
           carried out in the morphology-mass study 
           (top panel; Sect.~\ref{subsect:mass})
           and morphology-SFR study (bottom panel; Sect.~\ref{subsect:sfr}). 
           The metric
           of importance is the mean decrease in the residual sum of squares
           of the fit that results from splitting the data along the
           indicated axes. For the mass study, the most important predictor
           variables ($M_{20}$,$C$,$G$) are all associated with
           concentration, while for the SFR study the most important 
           variables ($C$,$G$,$I$) are associated with concentration
           as well as disturbance.
          }
  \label{fig:varimp}
\end{figure}

\begin{figure*}
  \includegraphics[width=6.5in]{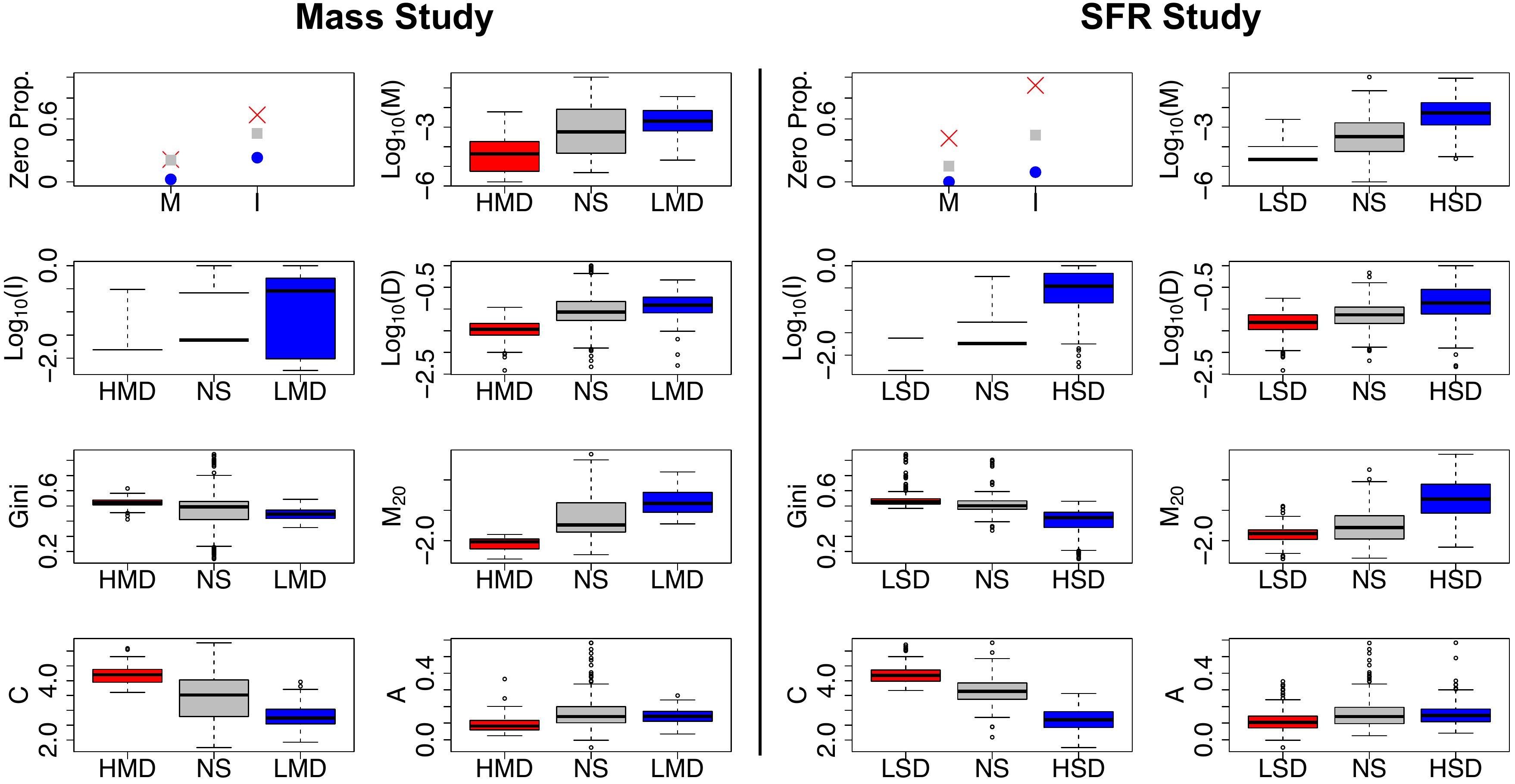}
  \caption{Boxplots summarizing the distributions of each morphological 
           statistic for the morphology-mass study (left bank of panels; 
           Sect.~\ref{subsect:mass}) and morphology-SFR study (right bank of
           panels; Sect.~\ref{subsect:sfr}). Within each bank, the leftmost
           boxplots (red) represent statistics for galaxies identified as being
           in high-mass-dominated (HMD) or low-SFR-dominated (LSD) regions,
           the rightmost boxplots (blue) are for galaxies in 
           low-mass-dominated (LMD) or high-SFR-dominated (HSD) regions, and
           the centre boxplots (grey), labeled ``NS" for ``not significant,"
           are for all other galaxies.
           The top left plot in each bank of panels indicates the
           proportion of zero values for the $M$ and $I$ statistics, with
           the same colour scheme as above. We observe that galaxies in
           HMD/LSD regions are more concentrated and less disturbed than their
           LMD/HSD region counterparts.
	   } 
  \label{fig:boxplot}
\end{figure*}

\subsection{Morphology-Mass Study}

\label{subsect:mass}

In this study, the predictor and response variables are respectively
\begin{itemize}
\item the morphological summary statistics $M$, $I$, $D$, $G$,
$M_{20}$, $C$, and $A$; and
\item the estimated stellar mass $M_{\ast} = \widehat{M}_{\rm med}$.
\end{itemize}
We sort the training set masses in ascending order and
identify the upper and lower quartiles (see the top panel of 
Fig.~\ref{fig:dists}). High- and low-mass galaxies are defined as those
with log$_{10}(M)$ $>$ 10.639 and $<$ 10.070 respectively, with the global
ratio of low- to high-mass galaxies being unity by definition.

The application of our local two-sample testing algorithm yields the
following results for the 700-galaxy test set:
\begin{itemize}
\item 108 galaxies are identified as lying in regions of the predictor space
where the proportion
of high-mass to low-mass galaxies is significantly larger than one,\footnote{
To be clear: a galaxy that is identified as lying in a region 
predominantly containing e.g.~high-mass galaxies is not necessarily itself a 
high-mass, upper-quartile galaxy.
}
while
\item 169 galaxies lie in primarily low-mass regions of predictor space.
\end{itemize}
In the left panel of Fig.~\ref{fig:varimp} we display variable importances
determined by the random forest.
These indicate that measures of light concentration
($M_{20}$, $C$, $G$) are more important than measures of disturbance 
($D$, $M$, $A$, $I$) for 
estimating local proportions of high- to low-mass galaxies. This observation
is consistent with the boxplots displayed in the left panels of 
Fig.~\ref{fig:boxplot}, which show the distributions of individual
morphological statistics for galaxies in high-mass-dominated (HMD; red) and
low-mass-dominated (LMD; blue) regions, as well as those for galaxies lying
where neither class dominates (grey). We observe that galaxies in 
HMD regions are more concentrated and less disturbed than their counterparts
in LMD regions, as one would expect given the redshift range of our sample
and previous results regarding ``cosmic downsizing"
(e.g.~\citealt{Cowie96};
see also Fig.~7 of \citealt{Bundy05}, which demonstrates at $z \sim 1$
that the fraction of galaxies classified as peculiar decreases as mass
increases$-$less disturbance$-$while the fraction of galaxies classified
as ellipticals increases with mass$-$greater concentration).

\begin{figure}
  \centering
  \includegraphics[width=2.75in]{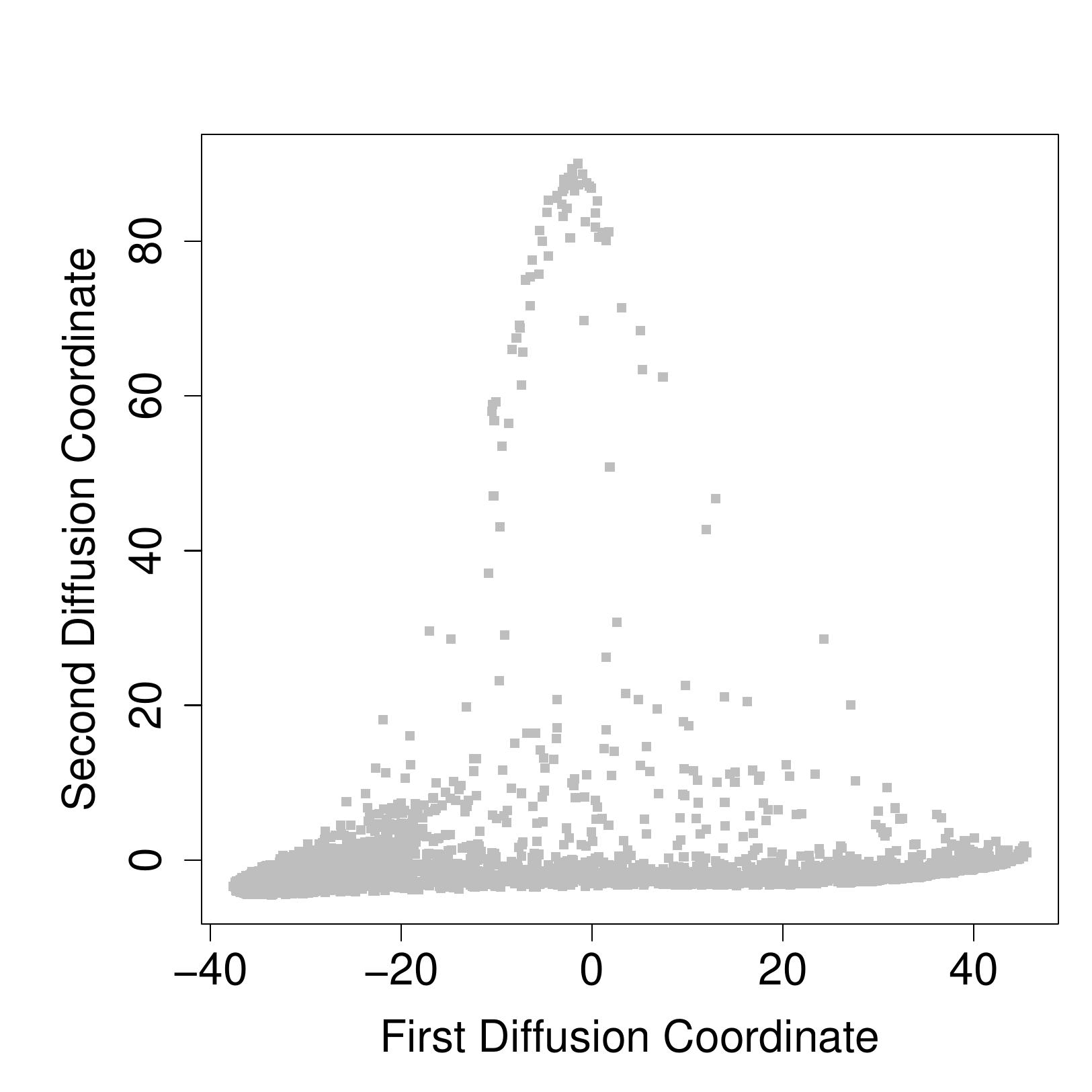}
  \caption{First two diffusion map coordinates resulting from the mapping
           of the seven-dimensional space of morphological statistics to
           diffusion space. Each point represents a single galaxy from the
           the full dataset. See Appendix \ref{app:dm} for more detail on how 
           one constructs a diffusion map. We stress that in our analysis the
           goal of using diffusion map is visualization: one should thus not 
           attempt to interpret the absolute numerical values along each 
           axis, which are not readily interpretable. Also, the reader should 
           keep in mind that two-sample testing is performed in the original 
           predictor space and not in diffusion space.
          }
  \label{fig:diffmap_all}
\end{figure}

To further visualize how the morphologies of galaxies, and their statistics,
change {\em within} HMD and LMD regions,
we utilize the diffusion map
algorithm (\citealt{Coifman05}, \citealt{Coifman06}, 
\citealt{Lafon06}, \citealt{Lee11}).
Diffusion map is useful for uncovering nonlinear
sparse structure in high-dimensional data, including submanifolds, clusters,
and high-density regions, etc. We constrast this with e.g.~principal
components analysis, a linear method wherein data are projected onto
hyperplanes. The interested reader can find further general details within the
astronomical literature in e.g.~\cite{Richards09} and \cite{Freeman09},
and details on our specific application of diffusion map in Appendix
\ref{app:dm}. The result of our application is to map galaxies from the
original seven-dimensional space of morphological statistics to an easily
visualized two-dimensional diffusion space. In Fig.~\ref{fig:diffmap_all},
we display the first two diffusion coordinates for all 2487 galaxies.

\begin{figure}
  \includegraphics[width=3.3in]{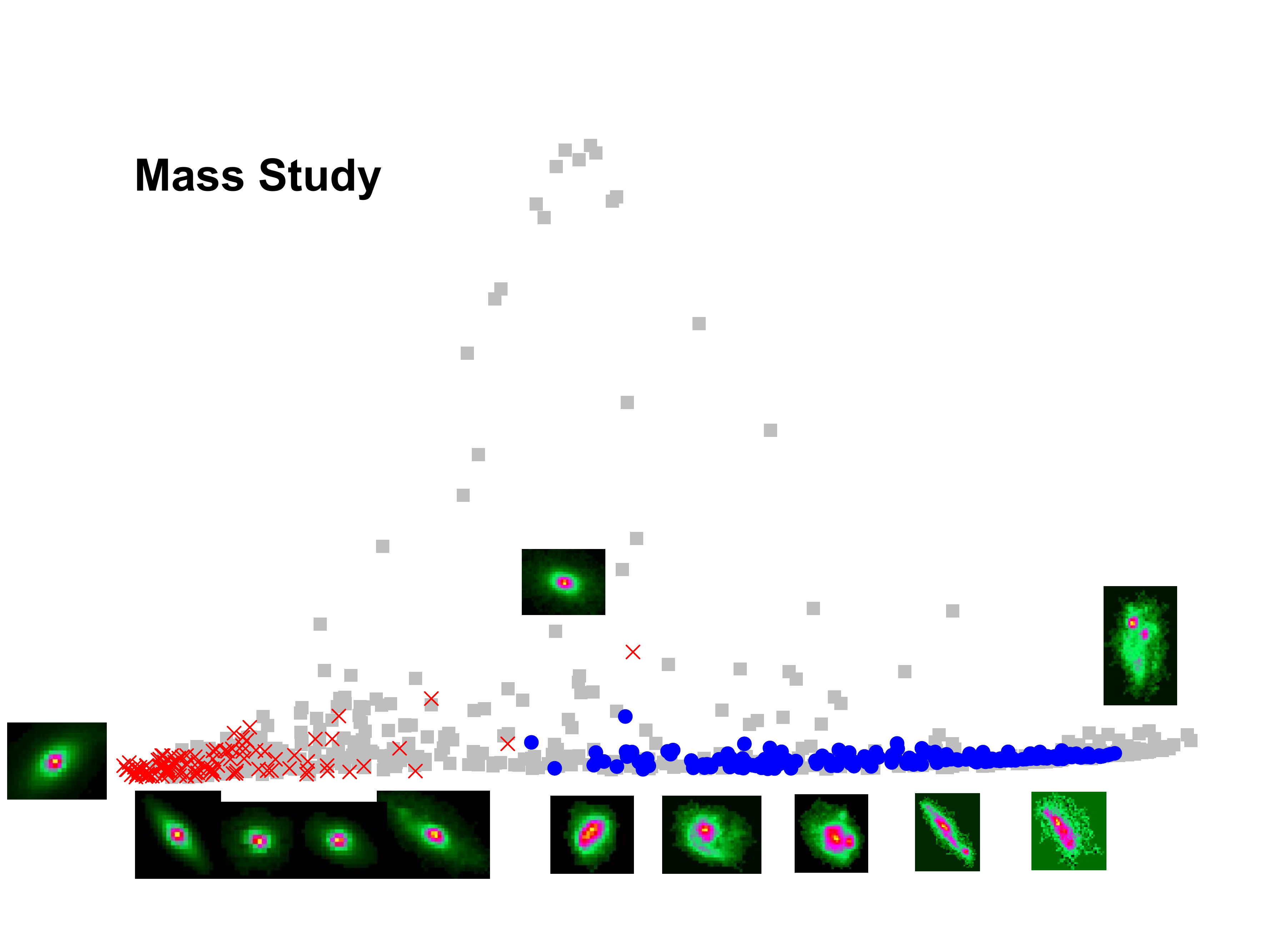}
  \includegraphics[width=3.3in]{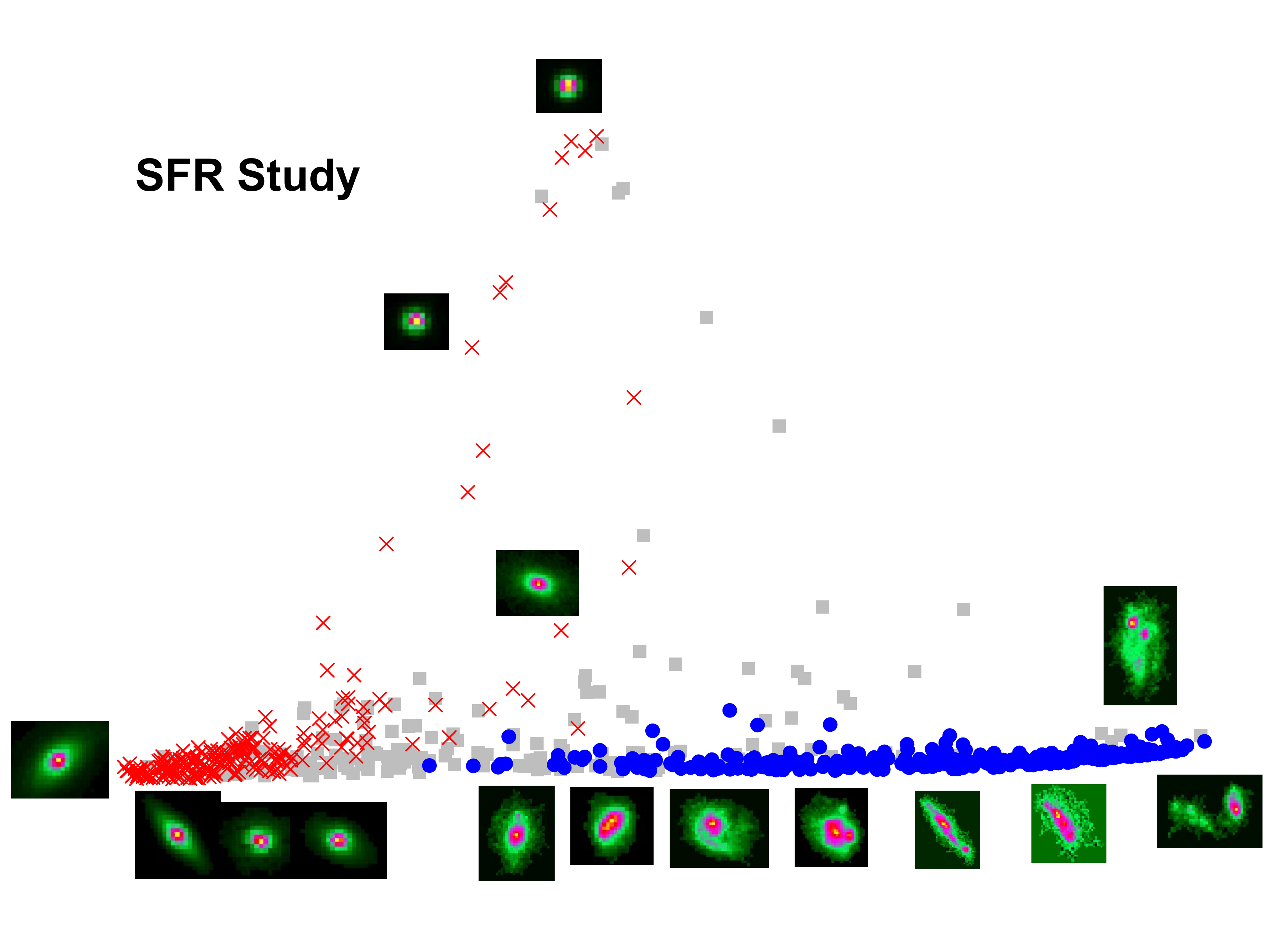}
  \caption{First two diffusion coordinates for the test data for the
           morphology-mass (top) and morphology-SFR (bottom) studies.
           Galaxies marked as red crosses are located in regions of the 
           seven-dimensional space of morphological statistics where the local
           proportion of high-mass (top) or low-SFR (bottom) galaxies is 
           significantly larger than the global proportion, whereas galaxies
           marked as blue circles reside in low-mass (top) or 
           high-SFR (bottom) regions.
           There is a clear separation between the regions along the
           first diffusion coordinate. Representative galaxies are 
           displayed for each group, near their locations in diffusion space.
           Galaxies to the left exhibit a consistent 
           appearance$-$concentrated, symmetric,
           undisturbed$-$while the galaxies to the right exhibit increasing
           amounts of the disturbance.
           }
  \label{fig:diffmap}
\end{figure}

\begin{figure*}
  \includegraphics[width=6.5in]{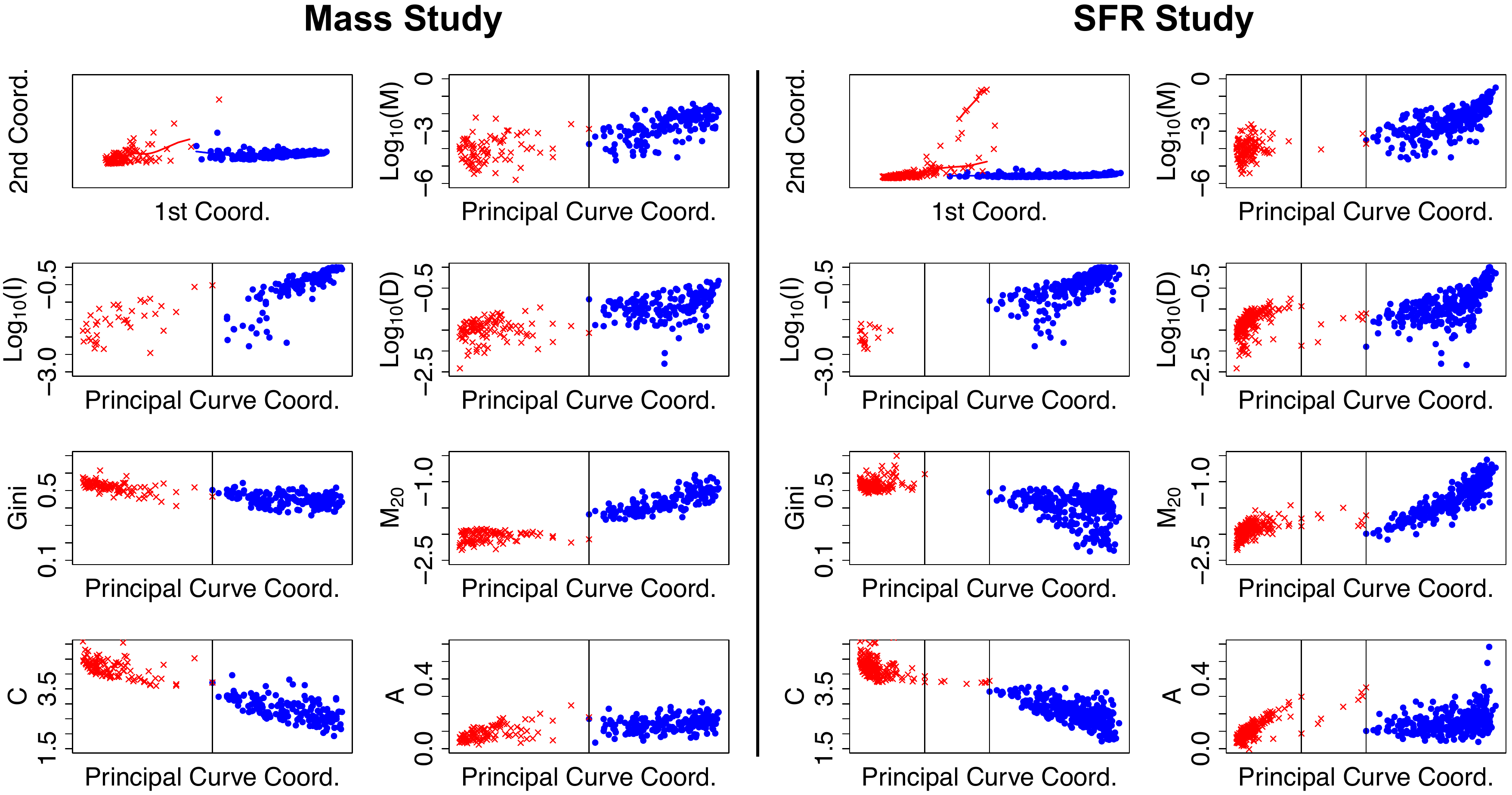}
  \caption{Scatter plots of each morphological statistic as a function of
           principal curve coordinate for the morphology-mass study 
           (left bank of panels; Sect.~\ref{subsect:mass}) and 
           morphology-SFR study (right bank of
           panels; Sect.~\ref{subsect:sfr}). These scatter plots take
           advantage of the enhanced data visualization afforded by 
           diffusion map to show
           more detail about the distributions of statistics than the
           summaries provided by the boxplots in Fig.~\ref{fig:boxplot}.
           The top left panel in each 
           bank shows the first two diffusion coordinates for galaxies
           identified as being in high-mass-dominated (HMD) or 
           low-SFR-dominated (LSD) regions (red crosses, towards the left
           in each panel), and 
           low-mass-dominated (LMD) or high-SFR-dominated (HSD) regions
           (blue circles, towards the right), 
           with principal curves for each set of points
           overlaid. (Note that the red and blue points are the same red
           and blue points displayed in Fig.~\ref{fig:diffmap}; we have
           simply removed the grey points displayed in that figure and
           decreased the plotting range for the second coordinate to
           enhance visibility.)
           In the remaining panels of each bank, the width of
           each box corresponds to the totality of each principal curve.
           Note that for the $M$ and $I$ panels, zero values are not
           displayed (leading to e.g.~the absence of points in the middle box 
           of the $I$ panel for the SFR study). The panels in both the
           left and right banks indicate that
           galaxies become progressively less concentrated and more disturbed
           as one move from left to right through the identified regions, 
           as well as across regions.
	   } 
  \label{fig:princurve}
\end{figure*}

In the top panel of Fig.~\ref{fig:diffmap}, we retain the diffusion 
coordinates for the test set data and mark those in HMD and LMD regions using
red crosses and blue circles, respectively.
We observe a clear separation HMD- and LMD-region galaxies
along the first diffusion map coordinate 
(i.e.~from left to right). Galaxies in the HMD region exhibit a consistent
appearance$-$concentrated, symmetric, undisturbed$-$with general evidence
of discs, while galaxies in the LMD region are generally less concentrated
and exhibit a range of appearances
due to increased disturbance that becomes more prevalent towards the right.
To further quantify these results, we plot the individual statistics of
each HMD and LMD region galaxy as a function of principal curve 
coordinates in the left bank of panels in Fig.~\ref{fig:princurve}.
Principal curves are smooth one-dimensional curves that provide a 
one-dimensional summary of multi-dimensional data (\citealt{Hastie89}).
In the top left panel of Fig.~\ref{fig:princurve}, the principal curves
that interpolate the HMD and LMD region galaxies are shown as solid red
and blue lines (to the left and right in each panel, respectively).
The other panels in the left bank indicate results 
consistent with those described above: galaxies become 
progressively less concentrated and more disturbed as one moves from left
to right within regions, and across regions.

\begin{figure}
  \centering
  \includegraphics[width=2.75in]{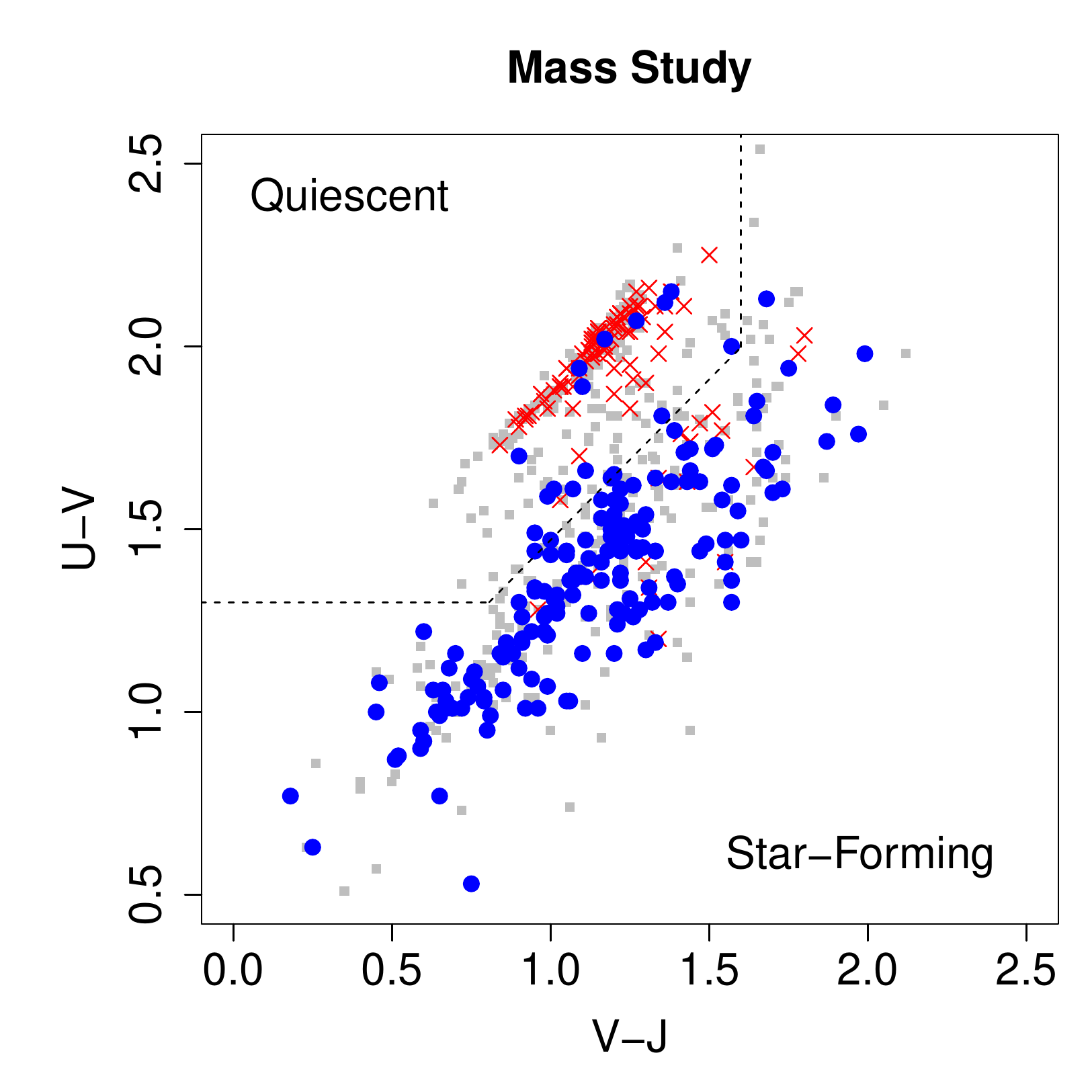}
  \includegraphics[width=2.75in]{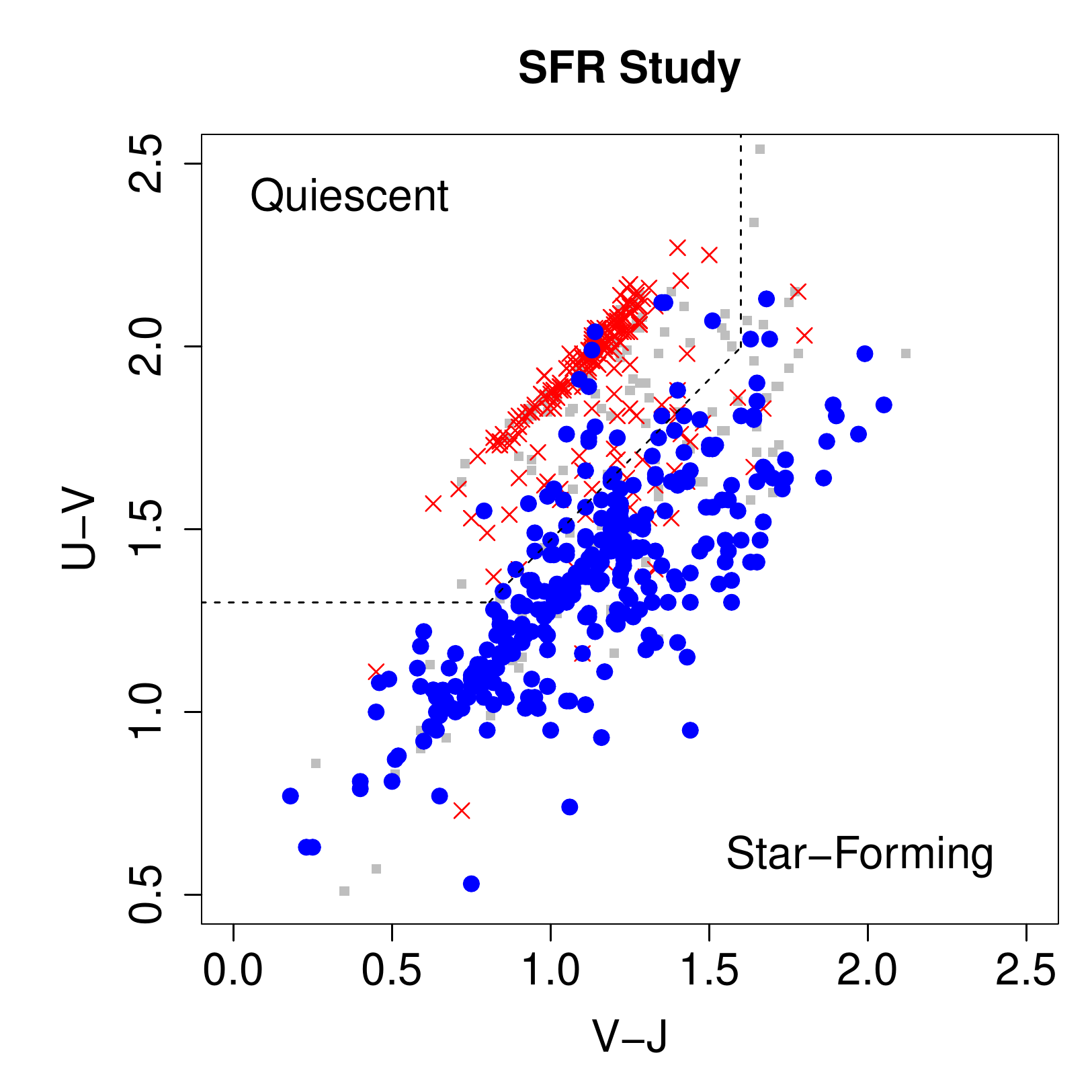}
  \caption{$UVJ$ diagrams for the morphology-mass study
           (top panel; Sect.~\ref{subsect:mass}) and 
           morphology-SFR study (bottom panel; Sect.~\ref{subsect:sfr}). 
           In the top panel, galaxies in HMD and LMD regions are 
           shown as red crosses and blue circles, respectively; in the
           bottom panel, LSD and HSD region galaxies are similarly shown.
           These diagnostic diagrams show that, consistent with a priori
           expectation, 
           quenched galaxies largely inhabit a tight locus on the $UVJ$
           diagram that lies above the less-tight locus of star-forming 
           galaxies.  
          }
  \label{fig:uvj_diag}
\end{figure}

\begin{figure}
  \centering
  \includegraphics[width=2.75in]{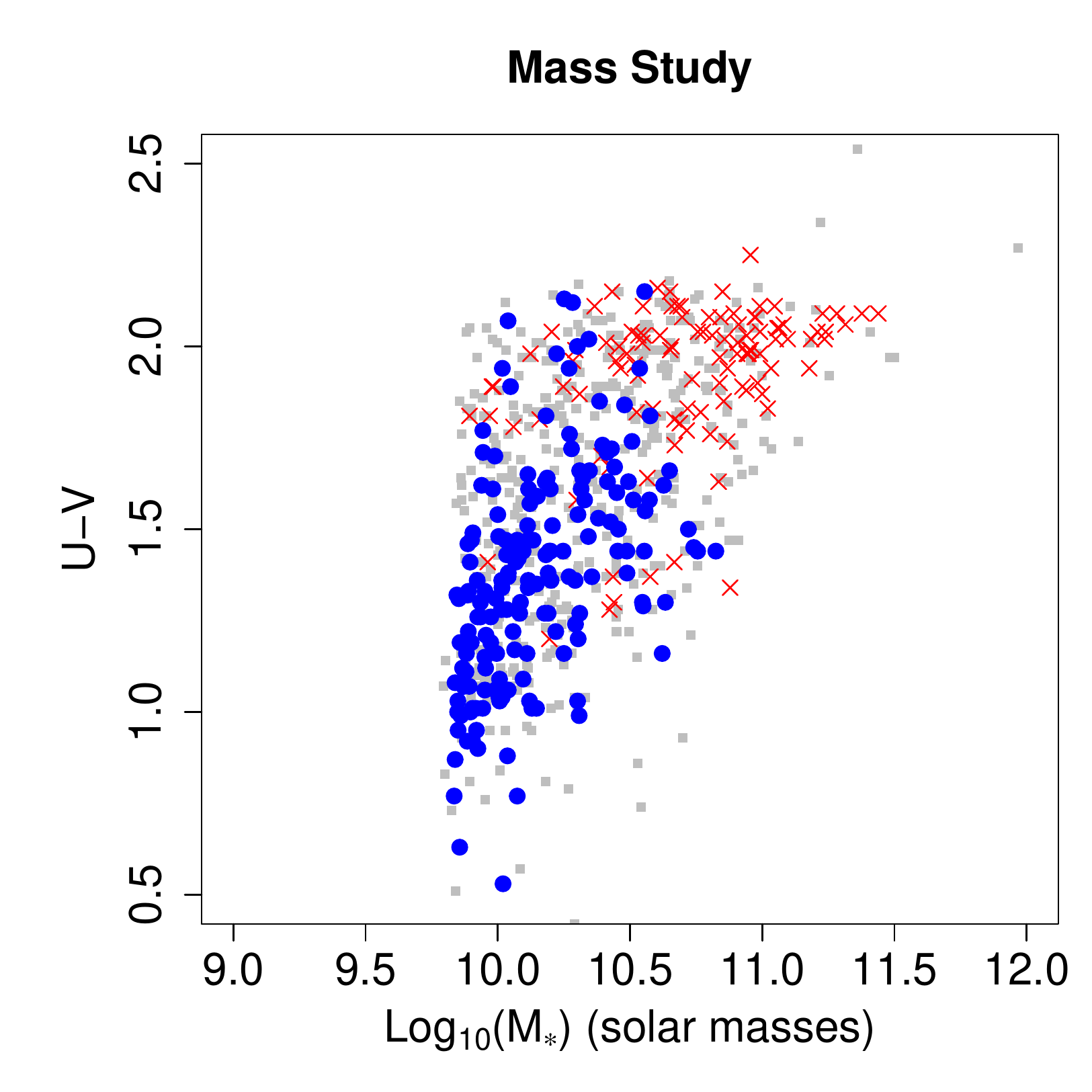}
  \includegraphics[width=2.75in]{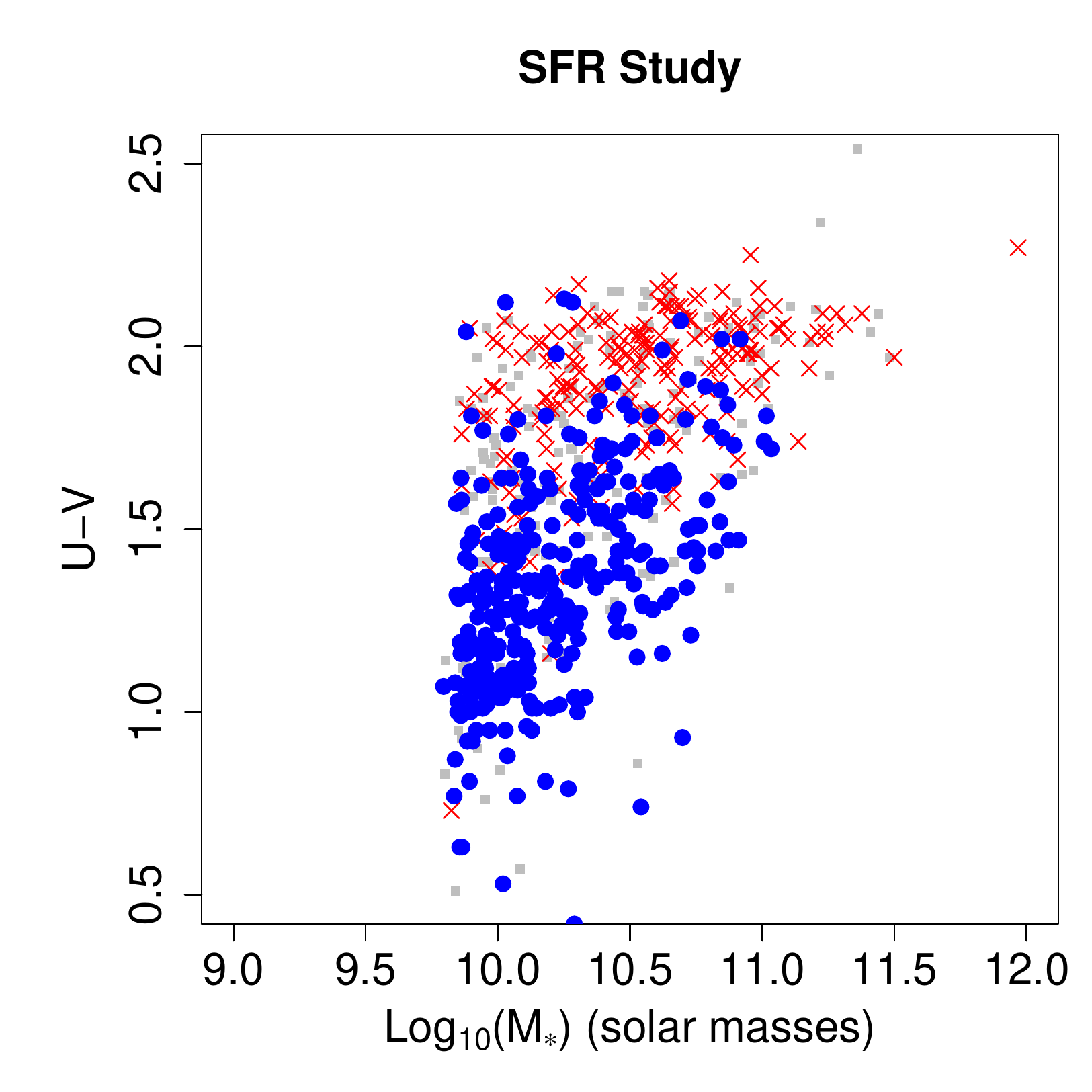}
  \caption{Same as Fig.~\ref{fig:uvj_diag}, except that here we plot $U-V$ 
           versus $M_{\ast}$. We observe that low-SFR (i.e.~quenched)
           galaxies are associated with higher mass galaxies.
          }
  \label{fig:uvmass_diag}
\end{figure}

To assess the consistency of our results with a priori expectation, 
we construct $UVJ$ and $M_{\ast}$-$UV$ diagrams 
(Figs.~\ref{fig:uvj_diag} and \ref{fig:uvmass_diag}).
In the top panel of Fig.~\ref{fig:uvj_diag},
quenched galaxies are identified as those that 
lie above the locus defined by $U - V = {\rm max}[1.3,0.88(V-J) + 0.59]$ and
the vertical line $V-J = 1.6$ (\citealt{Williams09}). We
observe that the majority of galaxies from HMD regions 
lie on the tight locus of quenched galaxies, as expected.
In the top panel of Fig.~\ref{fig:uvmass_diag}, 
quenched galaxies lie towards the top (with
$U-V$ values $\gtrsim$ 1.3). We observe a positive correlation between
being in a HMD region and being quenched, as expected. 

\subsection{Morphology-SFR Study}

\label{subsect:sfr}

In this study, the predictor and response variables are respectively
\begin{itemize}
\item the morphological summary statistics $M$, $I$, $D$, $G$,
$M_{20}$, $C$, and $A$; and
\item the estimated star formation rate $S$.
\end{itemize}
As is the case for the mass study, we sort the training set SFRs in
ascending order and define low- and high-SFR galaxies as those with
$\log_{10}(S) < -1.01$ and $> 1.232$ respectively. 
(See the bottom panel of Fig.~\ref{fig:dists}.)
The application of our two-sample testing algorithm yields that
\begin{itemize}
\item 313 of 700 test-set galaxies lie in high-SFR-dominated (HSD) regions, while another
\item 214 test-set galaxies lie in low-SFR-dominated (LSD) regions.
\end{itemize}
In the right panel of Fig.~\ref{fig:varimp}, we see that the most
important variables for estimating local proportions of high- to low-SFR
galaxies are those that measure the concentration of light ($C$ and $G$), 
although the separation between concentration-related and 
disturbance-related statistics is not as clear-cut as it is for the
mass study. As with the mass study, the boxplots displayed in the right
panels of Fig.~\ref{fig:boxplot} indicate clear differences between 
the distributions of summary statistics for each class of galaxy, with galaxies
in LSD regions being more concentrated and less disturbed than their
counterparts in HSD regions.\footnote{
We note here that we have also examined the specific star-formation rate in 
addition to the SFR; the results of our SSFR analysis are qualitatively 
similar and are not shown.
}

In the bottom panel of Fig.~\ref{fig:diffmap}, we display the first two
diffusion map coordinates for the galaxy test-set data, with galaxies in
LSD and HSD regions marked as red crosses and blue circles 
respectively. These regions are
largely but not entirely coincident with the HMD and LMD regions shown in
the top panel.\footnote{
See Table \ref{tab:assign}. We note that it (along with Fig.~\ref{fig:diffmap})
indicates that e.g.~to the left in the displayed diffusion map there are 
galaxies that
are associated with both the LSD and HMD regions, with just one or the other, 
or with neither region. 
We defer a detailed study of the morphological differences between
these four classes to a future work.} As is the case for the HMD
region, galaxies in the LSD region are concentrated, symmetric, and
undisturbed, but in addition to those galaxies that show visual evidence
of discs, there are disc-less galaxies, which cluster towards the top of
the panel. Conversely, as is the case for the LMD region, galaxies in the
HSD region are generally less concentrated and exhibit a range of disturbed
morphologies, particularly at the right end of the panel, a 
high-SFR-dominated but not low-mass-dominated region that is exclusively
populated by highly disturbed galaxies that are presumably undergoing mergers.
In the right bank of panels in Fig.~\ref{fig:princurve}, we show from left
to right the galaxy summary statistics for two principal curves fit to
LSD region galaxies and one fit to HSD region galaxies. Due to the 
near-coincidence of the LSD/HSD and HMD/LMD regions, the observed statistic
distributions are similar to those in the left bank of panels, but we do
note that the HSD region exhibits lower $G$, higher $M_{20}$, and higher
$A$ values, etc., than the LMD region, due to the fact that it contains
higher numbers of disturbed galaxies.

In Figs.~\ref{fig:uvj_diag} and \ref{fig:uvmass_diag}, we show
$UVJ$ and $M_{\ast}$-$UV$ diagrams with LSD and HSD regions highlighted.
As was the case for the mass study, we observe 
that the majority of galaxies in LSD regions lie on the locus of quenched 
galaxies, and that there is a positive correlation between
being in a LSD region and being quenched.

\begin{table}
\caption{Assignment of Test Data to Groups}
\begin{tabular}{lrrrr}
\hline
 & \multicolumn{1}{c}{Low SFR} & \multicolumn{1}{c}{Not Sig} & \multicolumn{1}{c}{High SFR} & \multicolumn{1}{c}{Total} \\
\hline
High Mass       &  78 &  30 &   0 & 108 \\
Not Sig         & 136 & 115 & 172 & 423 \\
Low Mass        &   0 &  28 & 141 & 169 \\
\hline
Total           & 214 & 173 & 313 & 700 \\
\hline
\label{tab:assign}
\end{tabular}
\end{table}

\section{Summary}

\label{sect:summary}

In this paper, we provide the astronomical community with a local
two-sample hypothesis test framework that one can use to more easily
analyse data (e.g.~morphological statistics, 
photometric magnitudes, mass and star-formation rate estimates, etc.)
in their native high-dimensional spaces.
In this framework, one defines two classes based on a response variable
of interest (e.g.~the top and bottom 25\% of the stellar masses for
a sample of galaxies), and uses regression to compute the class posterior
estimates $\widehat{\mathbb{P}}(Y=y \vert x)$
given a predictor datum $x$ and where $y$ denotes one of two discrete classes
(e.g.~high mass in a comparison of high mass and low mass, etc.).
We leverage the work of \cite{Wager14} and \cite{Wager15} to convert these
estimates to a asymptotically unbiased test statistic that under the
null hypothesis converges to a standard normal distribution
(eq.~\ref{eqn:test}).
In our implementation, we split data into training and test sets, using the
former to learn estimates $\widehat{\mathbb{P}}(Y=y \vert x)$ and
generating test statistics at the latter. To mitigate the effect of
multiple comparisons (i.e.~the fact that the number of tests performed
is greater than one), we apply the Benjamini-Hochberg procedure. More
details are provided in Algorithm \ref{alg:twosample} and {\tt R}-based
software is available at {\tt github.com/pefreeman/ltst}.

Our testing framework has a potential myriad of uses, as it is suitable for
use in any analysis situation in which one wishes to test whether a locally 
estimated 
proportion of two classes of objects is significantly different from the 
global proportion. In this paper, we demonstrate the efficacy of our
testing framework by 
applying it to a set of 2487 $i$-band-selected galaxies observed by
the HST ACS in the COSMOS, EGS, GOODS-North, and UDS fields. For these
galaxies, we compute seven morphological statistics ($M$, $I$, $D$, $G$,
$M_{20}$, $C$, $A$) and thus estimate $\mathbb{P}(Y=y \vert x)$ in
this seven-dimensional space. (We note that because our estimation makes use of
random forest regression, one can apply our framework to spaces of 
considerably higher dimensionality. For reference, the computation time
for our analyses are $\sim$ 1 CPU minute.)
We perform
two studies, one in which we determine the local proportion of high-mass
(top 25\% of masses) to low-mass (bottom 25\% of masses) galaxies, and 
another using star-formation rate in place of mass. Both studies yield
qualitatively similar results: galaxies lying in identified high-mass 
or low-SFR regions exhibit a consistent appearance$-$concentrated, 
symmetric, undisturbed, and generally with visual evidence of disc 
structure$-$while their counterparts in low-mass or high-SFR regions
have less concentrated light and exhibit increasing levels of disturbance.
We display these results first with boxplots (Fig.~\ref{fig:boxplot}) but
then show how one can further potentially visualize results at finer scales
by tranforming the predictor data to a lower-dimensional space; here,
we specifically apply diffusion map (Figs.~\ref{fig:diffmap} and
\ref{fig:princurve}). We provide details on diffusion map in
Appendix \ref{app:dm} and {\tt R}-based software that implements 
visualization via diffusion map at the address given above.

\section*{Acknowledgements}

The authors would like to thank the members of the CANDELS collaboration for
providing the data upon which this work is based. We would also like to thank
Jeff Newman (University of Pittsburgh) for acting as IK's external adviser 
for the project on which this
paper is based, and 
Rafael Izbicki (Federal University of S\~ao Carlos) and
Jen Lotz (Space Telescope Science Institute) for helpful
discussions. This work was supported by NSF DMS-1520786 and
NIMH R37MH057881. Our research has
made use of {\tt SAOimage DS9}, as well as
the {\tt dmtools} provided by the Chandra X-ray Center in the application
package {\it CIAO}.

\appendix

\section{Diffusion Map}

\label{app:dm}

Dimensionality reduction methods are useful for visualizing low-dimensional
structures embedded in higher-dimensional spaces. One such method is
diffusion map (\citealt{Coifman05}, \citealt{Lafon06}),\footnote{
Methods for computing diffusion coordinates, etc., are contained in
the {\tt R} package {\tt diffusionMap}.}
a nonlinear method that seeks to
preserve the connectivity structure of data within a high-dimensional
space. (In practice, preservation means that the Euclidean distance between two
points in diffusion space is approximately the same as the sum of all paths
between the same two points in the original data space.)
The connectivity structure is learned by modeling the traversal of the 
data space as a diffusion process.

As a starting point for constructing a diffusion map, one defines a weight
that reflects the local similarity of two points, $x_i$ and $x_j$, in 
$\cal{X}$ =
$\{x_1,\ldots,x_n\}$. In this work, we implement the weight estimator of
\cite{Zelnik05}:
\begin{equation}
\widehat{w}(x_i,x_j) = \exp\left(-\frac{s(x_i,x_j)^2}{\sigma_i\sigma_j}\right) \,,
\end{equation}
where $s$ is the (for example Euclidean) distance between $x_i$ and $x_j$,
and $\sigma_i$ ($\sigma_j$) is the distance between $x_i$ ($x_j$) and
its $k^{\rm th}$ nearest neighbour. (Note that we standardize the data of
each predictor variable first i.e.~from each datum we subtract the sample
mean and then divide the difference by the sample standard deviation.)
We assume $k = 30$; other values give
similar visualization results. (The appropriate value for $k$ will of
course differ from application to application.)

We use the weights $\widehat{w}$ to build a Markov random walk on the data
with the transition probability from $x_i$ to $x_j$ defined as
\begin{equation}
p(x_i,x_j) = \frac{\widehat{w}(x_i,x_j)}{\sum_{k=1}^n \widehat{w}(x_i,x_k)} \,.
\end{equation}
The one-step transition probabilities are stored in an $n \times n$ matrix
$\boldsymbol{P}$, and then propagated by a $t$-step Markov random walk with
transition probabilities $\boldsymbol{P}^t$. Instead of choosing a fixed
time parameter $t$, however, we combine diffusions at all times 
and define an averaged diffusion map\footnote{
Note this is the default way by which the {\tt diffusionMap} package function
{\tt diffuse()} constructs diffusion maps.}
according to
\begin{equation}
\Psi_{\rm av} : x \rightarrow \left[\left(\frac{\lambda_1}{1-\lambda_1}\right)\psi_1(x),\ldots,\left(\frac{\lambda_m}{1-\lambda_m}\right)\psi_m(x)\right] \,,
\end{equation}
where $\lambda_i$ and $\psi_i$ represent the first $m$ eigenvalues and 
right eigenvectors of $\boldsymbol{P}$.
In this work, we fix $m$ to 2, i.e.,
we only visualize the first two dimensions in diffusion space.

\bsp

\label{lastpage}


\begin{thebibliography}{}
\bibitem[Abraham et al.(1994)]{Abraham94} Abraham R.~G., Valdes F., Yee H.~K.~C., van den Bergh S., 1994, ApJ, 432, 75
\bibitem[Abraham et al.(1996a)]{Abraham96a} Abraham R.~G., Tanvir N.~R., Santiago B.~X., Ellis R.~S., Glazebrook K., van den Bergh S., 1996a, MNRAS, 279, 47L
\bibitem[Abraham et al.(1996b)]{Abraham96b} Abraham R.~G., van den Bergh S., Glazebrook K., Ellis R.~S., Santiago B.~X., Surma P., Griffiths R.~E., 1996b, ApJS, 107, 1
\bibitem[Abraham, van den Bergh \& Nair(2003)]{Abraham03} Abraham R.~G., van den Bergh S., Nair P., 2003, ApJ, 588, 218
\bibitem[Barro et al.(2014)]{Barro14} Barro G.~et al., 2014, ApJ, 791, 52
\bibitem[Behroozi, Wechsler \& Conroy(2013)]{Behroozi13} Behroozi P.~S., Wechsler R.~H., Conroy C.~2013, ApJ, 770, 57
\bibitem[Bell et al.(2012)]{Bell12} Bell E.~F.~et al., 2012, ApJ, 753, 167
\bibitem[Benjamini \& Hochberg(1995)]{Benjamini95} Benjamini Y., Hochberg Y., 1995, JRSS B, 57, 289
\bibitem[Bluck et al.(2014)]{Bluck14} Bluck A.~F.~L., Mendel J.~T., Ellison S.~L., Moreno J., Simard L., Patton D.~R., Starkenburg E., 2014, MNRAS, 441, 599
\bibitem[Bluck et al.(2016)]{Bluck16} Bluck A.~F.~L.~et al., 2016, MNRAS, 462, 2559
\bibitem[Brennan et al.(2017)]{Brennan17} Brennan R.~et al., 2017, MNRAS, 465, 619
\bibitem[Bundy, Ellis \& Conselice(2005)]{Bundy05} Bundy K., Ellis R.~S., Conselice C.~J., 2005, ApJ, 625, 621
\bibitem[Coifman \& Lafon(2006)]{Coifman06} Coifman R.~R., Lafon S., 2006, ACHA, 21, 5
\bibitem[Coifman et al.(2005)]{Coifman05} Coifman R.~R., Lafon S., Lee A.~B., Maggioni M., Nadler B., Warner F., Zucker S.~W., 2005, PNAS, 102, 7426
\bibitem[Conselice(2003)]{Conselice03} Conselice C.~J., 2003, ApJS, 147, 1
\bibitem[Conselice(2014)]{Conselice14} Conselice C.~J., 2014, ARA\&A, 52, 291
\bibitem[Cowie et al.(1996)]{Cowie96} Cowie L.~L., Songaila A., Hu E.~M., Cohen J.~G., 1996, AJ, 112, 839
\bibitem[Dahlen et al.(2013)]{Dahlen13} Dahlen T.~et al., 2013, ApJ, 775, 93
\bibitem[Duong(2013)]{Duong13} Duong T., 2013, JNS, 25, 635
\bibitem[Elbaz et al.(2011)]{Elbaz11} Elbaz D.~et al., 2011, A\&A, 533, A119
\bibitem[Fang et al.(2015)]{Fang15} Fang G., Ma Z., Kong X., Fan L., 2015, ApJ, 807, 139
\bibitem[Freeman et al.(2009)]{Freeman09} Freeman P.~E., Newman J.~A., Lee A.~B., Richards J.~W., Schafer C.~M., 2009, MNRAS, 398, 2012
\bibitem[Freeman et al.(2013)]{Freeman13} Freeman P.~E., Izbicki R., Lee A.~B., Newman J.~A., Conselice C.~J., Koekemoer A.~M., Lotz J.~M., Mozena M., 2013, MNRAS, 434, 282
\bibitem[Gretton et al.(2012)]{Gretton12} Gretton A., Borgwardt K., Rasch M., Schoelkopf B., Smola A., 2012, JMLR, 13, 723
\bibitem[Grogin et al.(2011)]{Grogin11} Grogin N.~A.~et al., 2011, ApJS, 197, 35
\bibitem[Hastie \& Stuetzle(1989)]{Hastie89} Hastie T., Stuetzle W., 1989, JASA, 84, 502
\bibitem[Huertas-Company et al.(2016)]{Huertas16} Huertas-Company M.~et al., 2016, MNRAS, 462, 4495
\bibitem[Koekemoer et al.(2011)]{Koekemoer11} Koekemoer A.~M.~et al., 2011, ApJS, 197, 36
%\bibitem[James et al.(2013)]{James13} James G., Witten D., Hastie T., Tibshirani R., An Introduction to Statistical Learning with Applications in R, Springer, Berlin
\bibitem[Lackner \& Gunn(2013)]{Lackner13} Lackner C.~N., Gunn J.~E., 2013, MNRAS, 428, 2141
\bibitem[Lafon \& Lee(2006)]{Lafon06} Lafon S., Lee A.~B., 2006, IEEE Trans. Pattern Analysis and Machine Intelligence, 28, 1393
\bibitem[Lang et al.(2014)]{Lang14} Lang P.~et al., 2014, ApJ, 788, 11
\bibitem[Lee \& Freeman(2011)]{Lee11} Lee A.~B., Freeman P.~E., 2012, in Feigelson E., Babu G., eds., Statistical Challenges in Modern Astronomy V, Springer, New York, p.~255
\bibitem[Lotz, Primack \& Madau(2004)]{Lotz04} Lotz J.~M., Primack J., Madau P., 2004, AJ, 128, 163
\bibitem[Mobasher et al.(2015)]{Mobasher15} Mobasher B.~et al., 2015, ApJ, 808, 101
%\bibitem[Murtagh \& Legendre(2014)]{Murtagh14} Murtagh F., Legendre P., 2014, JClass, 31, 274
\bibitem[Peth et al.(2016)]{Peth16} Peth M.~A.~et al., 2016, MNRAS, 458, 963
\bibitem[Richards et al.(2009)]{Richards09} Richards J.~W., Freeman P.~E., Lee A.~B., Schafer C.~M., 2009, ApJ, 691, 32
\bibitem[Roederer \& Hardy(2001)]{Roederer01} Roederer M., Hardy R.~R., 2001, Cytometry, 45, 56
\bibitem[Salmi et al.(2012)]{Salmi12} Salmi F., Daddi E., Elbaz D., Sargent M.~T., Dickinson M., Renzini A., Bethermin M., Le Borgne D., 2012, ApJ, 754, L14
\bibitem[Santini et al.(2015)]{Santini15} Santini P.~et al., 2015, ApJ, 801, 97
\bibitem[Schaye et al.(2015)]{Schaye15} Schaye J.~et al., 2015, MNRAS, 446, 521
\bibitem[S\'ersic(1963)]{Sersic63} S\'ersic J.~L., 1963, BAAA, 6, 41
\bibitem[Simmons et al.(2017)]{Simmons17} Simmons B.~et al., 2017, MNRAS, 464, 4420
\bibitem[Snyder et al.(2015)]{Snyder15} Snyder G.~F.~et al., 2015, 454, 1886
\bibitem[Sz\'ekely \& Rizzo(2004)]{Szekely04} Sz\'ekely G.~J., Rizzo M.~L., 2004, InterStat, 5, 1
\bibitem[Teimoorinia, Bluck \& Ellison(2016)]{Teimoorinia16} Teimoorinia H., Bluck A.~F.~L, Ellison S.~L., 2016, MNRAS, 457, 2086
\bibitem[Vogelsberger et al.(2014)]{Vogelsberger14} Vogelsberger M.~et al., 2014, MNRAS, 444, 1518
\bibitem[Wager \& Athey(2015)]{Wager15} Wager S., Athey S., 2015, arXiv:1510.04342
\bibitem[Wager, Hastie \& Efron(2014)]{Wager14} Wager S., Hastie T., Efron B., 2014, JMLR, 15, 1625
\bibitem[Wake, van Dokkum \& Franx(2012)]{Wake12} Wake D.~A., van Dokkum P.~G., Franx M., 2012, ApJ, 751, L44
\bibitem[Weyant, Schafer \& Wood-Vasey(2013)]{Weyant13} Weyant A., Schafer C., Wood-Vasey W.~M., 2013, ApJ, 764, 116
\bibitem[Williams et al.(2009)]{Williams09} Williams R.~J., Quadri R.~F., Franx M., van Dokkum P., Labb\'e I., 2009, ApJ, 691, 1879
\bibitem[Woo et al.(2015)]{Woo15} Woo J., Dekel A., Faber S.~M., Koo D.~C., 2015, MNRAS, 448, 237
\bibitem[Wuyts et al.(2011)]{Wuyts11} Wuyts S.~et al., 2011, ApJ, 742, 96
\bibitem[Zelnik-Manor \& Perona(2005)]{Zelnik05} Zelnik-Manor L., Perona P.~2005, ANIPS, 17, 1601
\end{thebibliography}
\end{document}